# Beta-Borophene Under Circularly Polarized Radiation: Polaritonic and Polaronic Dynamic Band Structure

D. Akay[1], J. Schliemann[2]

[1]Department of Physics, Faculty of Science, Ankara University, Tandogan Ankara 06100, Turkey

[2]Institute for Theoretical Physics, University of Regensburg, Regensburg, Germany

**Abstract**

We study the effect of circularly polarized electromagnetic radiation on optical polarons in monolayer $\beta$-borophene. We focus on the off-resonant regime of large driving frequencies, which allows to set up an effectively time-independent Hamiltonian describing the radiation field in up to second perturbational order. The contributions of optical surface phonons are incorporated according to Lee-Low-Pines theory, via a sequence of appropriate unitary transformations. A central object in our investigation is the electronic band gap, which is shown to be widely controllable via the variation of light irradiation and substrate properties.

1. **Introduction**

Since the exfoliation of graphene [1], two-dimensional (2D) materials have been widely studied [2−4]. Many noteworthy physical characteristics of 2D materials make them attractive for use in photonics, energy transportation, nanoengineering, and electronic devices [5−9]. These unique Dirac materials exhibit linear energy dispersion in the Brillouin zone and provide an excellent platform for investigating quantum topological states. These characteristics of which are preserved by a topological invariant [10−14]. The 2D material system may also be used as a novel tool to detect Berry curvature in superlattice effects in twisted 2D materials [15]. A new series of thin atomic materials with exceptional characteristics is borophene (2D boron sheet), a revolutionary 2D material of group-III. It has received much interest because of its amazing physical and chemical characteristics [16−18]. The discovery of the structure's amazing characteristics has triggered much new research. Until 2015, boron-based atom-thick materials have been implemented experimentally on metal substrates [19−26]. The 2D boron systems are known to be highly dependent on substrate-induced effects [27]. Until this date, only a few synthetic borophene sheets have been empirically established. Although borophene may be synthesized efficiently and effectively, it is uncertain whether synthetic borophene can exist in structurally and chemically distinct layers or not.

Borophene is a neighbor of graphene, therefore it should have certain characteristics comparable to graphene [28−31]. Borophene has both $\sigma$ and $\pi$ electrons that occupy the electronic states of the Fermi surface. In particular, this material has been predicted to exhibit phonon-mediated superconductivity [32−34]. Nevertheless, the theoretically anticipated and produced planar structure differs from the other 2D materials because of the various out-of-plane buckling degrees. Furthermore, investigating the honeycomb borophene has appealing reasons, such as the Dirac fermions in the honeycomb lattice having similar electrical characteristics to Group-IV (monoelemental-enes) two-dimensional materials [35,36]. Recently, several studies have extensively investigated that the electron deficit of boron makes honeycomb borophene difficult to manufacture [37], but the unstable nature allows it to meet the biodegradability requirement for biomedical applications. The mechanical characteristics of borophene are particularly interesting: (i) Borophene has a low mass density and is the lightest 2D material known to date. Therefore, borophene can be used as an aide element in the construction of composites if its ideal strength and in-plane stiffness are sufficiently high. (ii) Because of its great flexibility against off-plane deformation, borophene is well suited for the fabrication of flexible nanodevices [38−40].

Furthermore, borophene's lattice thermal conductivity is extremely anisotropic, and its magnetic and electrical characteristics can be efficiently tuned for various applications. Moreover, borophene has significant potential as a material for touch screen technologies due to its high Fermi velocity, which is four times that of graphene, strong metallicity, and excellent optical transparency [41−43]. Borophene also has a strong ionic conductivity and great electronic conductivity. It performs rather well throughout the charging and discharging process. The transmittance of borophene is high along the uncorrupted direction, so the electrical conductivity and ionic conductivity are quite high. However, borophene has a relatively low optical conductivity in the visible range. Such characteristics open up possibilities for use in flexible electronics, photovoltaics, and display technologies. Furthermore, following oxidation, borophene remains metallic, and oxidized borophene improves significantly in both optical conductivity and properties [44].

Two-dimensional (2D) materials' electronic states and characteristics might be physically manipulated and even fine adjusted successfully by adopting so-called Floquet engineering, which involves applying an off-resonant and high-frequency treatment field with varied polarizations. The practical application of such a semiclassical dressing technique with a nonionizing but powerful laser field has only recently been conceivable due to major advancements in the microwave, laser, and terahertz technology. The alteration of electronic characteristics due to external irradiation has been extensively studied theoretically using the Floquet theory [45−48] for periodically driven quantum systems [49], which cover a wide spectrum of 2D materials. The authors of Ref. [50] to determine the Drude conductivity in the presence of a non-perturbative external driving, unify linear response theory and Floquet formalism. In the case of indirect bandgaps, an extra phonon must be emitted or absorbed to compensate for the energy imbalance, reducing the efficiency of the photon emission or absorption process [51]. Tai et al. determined the optical bandgap of boron to be 2.25 eV [52]. I get close to the figure confirmed by first-principles calculations, which was 2.07 eV [52]. They discovered that borophene is a beautiful direct bandgap semiconductor for intense photoluminescence [52]. Borophene is highly photosensitive to surface changes, as demonstrated by Lherbier et al. [53]. Furthermore, some publications have indicated that borophenes are poor absorbers in the visible region, resulting in good optical transparency in any direction [54,55]. The capacity of semiconductors to emit and absorb light is directly controlled by their electrical band structures. The low mass density of boron causes electron-phonon combinations to solidify, which might increase phonon-mediated superconductivity with high critical temperature [56,57]. Boron displays a lot more crossings in its bands, which are quite scattered (nearly parabolic). This shows that 2D boron has a high free charge shipper concentration as compared to metallic MX2 compounds [58]. As a result, borophene might be a promising future electrode material [53]. In summary, given the increased interest in this material, a precise model explaining the band structure and electrical characteristics is extremely desirable. The application of light to the surface of graphene produces a photoinduced mass term because of the time-reversal symmetry being broken [29]. But this model, as an alternative to the traditional technique, has the benefit of allowing for a great degree of controllability of perturbation parameters. There have also been significant theoretical and practical efforts recently to design the topological features of 2D Dirac systems using off-resonant light [59−61].

The purpose of this paper is to understand the role of circularly polarized beams in the mini band gap induced by the interaction of charge carriers $\beta$-borophene with surface optical phonon modes. For this, a circularly polarized beam on monolayer borophene was written with Floquet-Magnus perturbative expansion for an off-resonant and high-frequency periodic dressing field. Then, inside the continuum limit, we construct a Fröhlich-type Hamiltonian that includes both electron–SO phonon and electron-photon interactions, and to diagonalize the associated Hamiltonian, we use a unitary transformation scheme based on Lee–Low–Pines (LLP) theory [62]. The energy spectrum of electrons interacting with SO phonons in the irradiated monolayer-borophene is derived analytically within the context of this theory.

The following is how this paper is structured. In the next section, we will introduce our model, as well as its findings and debates. We conclude with a brief Conclusion section.

## 2. Model Hamiltonian

The low-energy effective Hamiltonian for $\beta$-Pmmn borophene with tilted anisotropic Dirac cones is as follows [63]:

$$H = \eta\hbar(v_x k_x \sigma_x + v_y k_y \sigma_y + v_t k_y \sigma_0) \quad (1)$$

where $\eta$ indicates $\pm 1$ valleys, $\sigma_i$ ) are the Pauli matrices, and $k = (k_x, k_y)$ are the x and y in-plane wavevectors. $v_x$ and $v_y$ are the velocities of Dirac electrons traveling along the $x$ and $y$-axes, respectively, and $v_t$ is the tilted velocity. Their magnitudes are in the range of, $0.86 v_F, 0.69 v_F$, and $0.32 v_F$, with $v_F = 10^6\, m/s$ [63] and their velocity depends on the bond length of the borophene system, the nearest neighbor hopping energy of the electrons. The Hamiltonian's Eq. (1) energy dispersion is calculated as,

$$E = \eta\hbar v_t k_y + \lambda\hbar\sqrt{v_x^2 k_x^2 + v_y^2 k_y^2} \quad (2)$$

It is made up of a Dirac cone for holes and electrons that touch each other at $k = 0$ like graphene. However, in this situation, they are deformed along the $k_y$ axis. The Dirac cone in valley $\eta = -1$ is slanted to the negative axis, whereas it is inclined to the positive axis in valley $\eta = 1$. $\lambda$ is the chirality index, which defines the energy branch and takes $1(-1)$ values corresponding to conduction (valence) bands in each valley. Borophene's Fermi surface differs from graphene's, which has an isotropic circular form with a radius $E_F/\hbar v_F$. However, the Fermi surface for borophene is elliptical due to the anisotropy of the Fermi velocities [57]. Consider an off-resonant circularly polarized light beam (CPL). This CPL may be represented mathematically by creating a time-dependent vector potential $\vec{A}$ defined as, $\vec{A} = A_0(\xi \sin(\Omega t)\hat{x} + \cos(\Omega t)\hat{y})$. $A_0$ is the off-resonant CPL of amplitude and $\Omega$ is the frequency of CPL. $A_0 \Omega$ is the field strength of the electric field. Here, $\xi = 1(-1)$ defines left (right) handed for circularly polarized light. The light source is a periodic external electromagnetic disturbance that serves to gap the Dirac dispersion, as in the traditional model that induced Haldane mass [29].

By Peierls substitution, i.e., $\vec{k} \rightarrow \vec{k} + \vec{A}(t)$ the time-dependent vector potential can be incorporated into the anisotropic Dirac Hamiltonian Eq.(1). The Hamiltonian, satisfying a time-dependent Hamiltonian of the form $\hat{H}(\vec{k}, t) = \hat{H}(\vec{k}, t + T)$ with period time $T = 2\pi/\Omega$. This periodic Hamiltonian can be extended into a Fourier series $H(\vec{k}, t) = \sum_n H_n(\vec{k}) e^{in\Omega t}$ using the Floquet theory. The zero-photon state may be separated from the other states and just analyze its dressed state influence through virtual photon absorption/emission processes [64,65] in the off-resonant limit $A_0^2/\Omega \ll 1$. This results in an effective static Hamiltonian being used to characterize the dressed system.

Noted that the radiation states can contain any number of photons in different modes. Nonetheless, we concentrate on a process in which a single photon is emitted (absorbed), leaving the other photons as passive observers with no impact on the process's amplitude. Thus, for a light-matter interaction involving only a single photon, the Floquet-Magnus expansion technique can be used to change to a time-independent effective Hamiltonian [66]. It is used to block and diagonalize the entire Floquet Hamiltonian perturbatively. Here, we only consider the $n = 0$ and $1\ subspaces$, decoupling the higher-order Floquet states from the zero-photon state ($n = 0$). Also known as the Floquet eigenvalue problem,

$$\sum_{m,n \epsilon Z}(H_{mn} - m\Omega \delta_{mn})|\phi_n\rangle = \varepsilon_k |\phi_n\rangle. \quad (3)$$

Integrated partially by perturbative corrections, ignoring small, negligible off-diagonal blocks written up to second order perturbation as $H_\eta^{eff} \cong H_0 + H^1 + H^2$. The expression is near the Dirac point [64]:

$$H_\eta^{eff}(\vec{k}) \approx H_0(\vec{k}) + \sum_{n \neq 0} \frac{1}{n\Omega}[H_{-n}(\vec{k}), H_{+n}(\vec{k})] + O\left(\frac{1}{\Omega^2}\right) \quad (4)$$

In here, $H_n$ can be described by, $H_n = \frac{1}{T}\int_0^T dt e^{\Omega t} H_t$, which is the nth Fourier harmonic of the time-periodic Hamiltonian. $H_t$ is time dependent part. The Fourier components are calculated using the Hamiltonian in Eq.(1). Changes into a time-dependent form by applying the Peierls substitution $\vec{k} \to \vec{k} + \vec{A}(t)$ representing the coupling of the electromagnetic field to the Hamiltonian. The formula can be written as [67],

$$H_\eta^{eff}(\vec{k}) \approx H_0(k) + \frac{1}{\hbar\Omega}[H_{-1}(\vec{k}), H_1(\vec{k})] + \frac{1}{2(\hbar\Omega)^2}\{[[H_1(\vec{k}), H_0(\vec{k})], H_{-1}(\vec{k})] + H.C.\} \quad (5)$$

$H_0$ describes a system where no photon exchange occurs, while $H_1(H_{-1})$ considers the emission (absorption) of one virtual photon. In the case of the polarization of the linearly polarized off-resonant light along x-direction renormalized velocities $v_x^{eff} = (1 - \alpha_x)v_x$ together with $\alpha_x = \left(\frac{eA_0 v_y}{\hbar\Omega}\right)^2$. In the same way, in the case of the polarization of the linearly polarized off-resonant light along y-direction renormalized velocities $v_y^{eff} = (1 - \alpha_y)v_y$, with $\alpha_y = \left(\frac{eA_0 v_x}{\hbar\Omega}\right)^2$. After obtaining the renormalized velocities, the effective time-independent Hamiltonian part serves as the starting point for examining models of light-matter interaction. We proceed to construct the time-independent renormalized Dirac Hamiltonian via renormalized velocities, which has the form

$$H_\eta^{eff}(\vec{k}) \approx \eta\hbar[v_x^{eff} k_x \sigma_x + v_y^{eff} k_y \sigma_y + v_t k_y \sigma_0] + \Delta\sigma_z. \quad (6)$$

where $\Delta = \xi e^2 A_0^2 v_x v_y/\hbar\Omega$ is the effective energy term describing the effects of the circularly polarized light, which essentially renormalized the mass of the Dirac fermions. In Eq.(6) $\sigma_i (i = x, y, z)$ is Pauli matrices. Take note that equation (4) is only valid, if the photon energy is considerably larger than the band gap, $\hbar\Omega \gg t$ (t is proportional to the bandwidth) with $t = v\hbar/\sqrt{2}\tilde{a}$. Because, perturbation theories are applicable $\tilde{a}$ is the lattice constant and its value is approximately 1.62Å. The electric field corresponding to the vector potential used in, $\vec{A}(t) = A_0(\xi sin(\Omega t)\hat{x} + cos(\Omega t)\hat{y})$, $A_0$ is the amplitude and $\Omega$ is the frequency of associated off-resonant light. Besides, $A_0 = E/\Omega$ with E is the amplitude of the electric field $\vec{E}(t) = \partial\vec{A}(t)/\partial t$. With the average one period, the intensity can be written as,

$$I = \frac{cn\varepsilon_0}{2}(A_0\Omega)^2 \quad (7)$$

$c$ is the speed of light, $\varepsilon_0$ is the dielectric constant, and $n$ is the refractive index. The driving frequency is assumed to be in the laser frequency region. Considering these definitions, the effective energy term describing the effects of circularly polarized light in terms of the intensity can be calculated,

$$\Delta = \frac{\xi 2e^2 I v_x v_y}{c\varepsilon_0 \hbar\Omega^3} \quad (8)$$

The rearrangement of bands via the additional Floquet term introduced $\Delta$ gap term, i.e., mass term. The gapping of the massless Dirac band manifests as an anomalous Hall effect (AHE)-driven currents [68,69]. Anomalous Hall currents created by circularly polarized light are predicted to have some characteristics which directly detect the difference between effective gaps generated with right-left circular polarization ($\Delta[\circlearrowright - \circlearrowleft]I$). This can be realized, e.g. graphene by using an optical polarization chopping technique [69]. Thus, we can derive driven currents that represented the mass term and which can be evaluated as a photoinduced (PI) mass. In this study, we have used right-handed for circularly polarized light and the PI mass term is acquired by electrons via virtual absorption of off-resonant photon energy. Thus, the corresponding energy dispersion relation can be obtained as,

$$E_0 = \eta\hbar v_t k_y + \lambda\hbar\sqrt{\left(v_x^{eff} k_x\right)^2 + \left(v_y^{eff} k_y\right)^2 + \frac{\Delta^2}{\hbar^2}} \quad (9)$$

which can be written as,

$$E_0 = \eta\hbar v_t k_y + \lambda E_{CV}$$

Here, $\eta$ describes the valley index and $\lambda$ is the band index, which takes $+1(-1)$ the sign is for the conduction (valence) band in $\beta$-borophene.

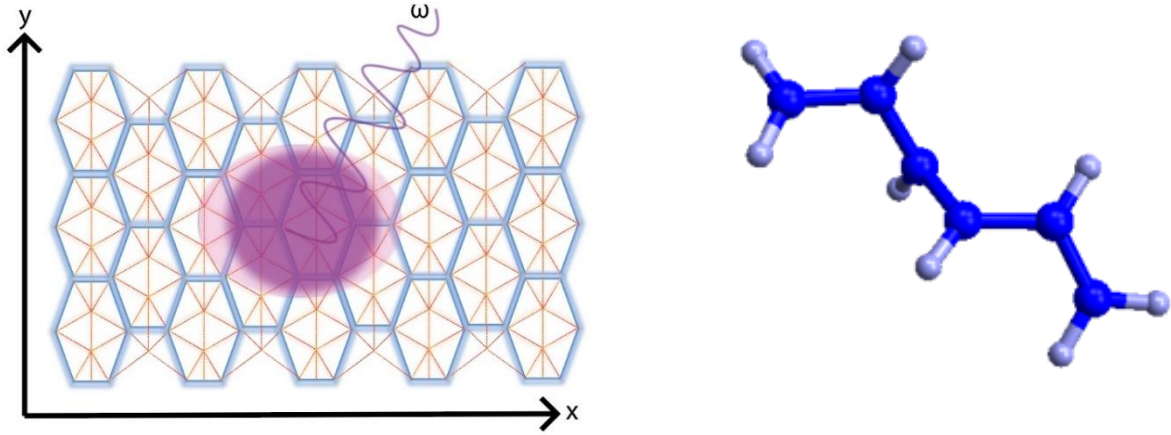

**Fig.1 (a)** Top view of the optimized borophene crystal structure (virtual photon absorption/emission processes), **(b)** 3D schematic representation of the borophene lattice.

Calculated eigenfunctions can be constructed in terms of pseudospinors with the associated two-component eigenvectors

$$\Psi = \frac{1}{\sqrt{2}}\begin{pmatrix} \sqrt{1+\frac{\Delta}{E_{CV}}}\,e^{i\phi} \\ \sqrt{1-\frac{\Delta}{E_{CV}}} \end{pmatrix} \quad (10)$$

where, $E_{CV}(k_x, k_y) = \hbar\sqrt{\left(v_x^{eff} k_x\right)^2 + \left(v_y^{eff} k_y\right)^2 + \frac{\Delta^2}{\hbar^2}}$ and $\phi = arctan\left(\frac{v_y^{eff} k_y}{v_x^{eff} k_x}\right)$. The electron Hamiltonian of Eq.(6) represents the system when a borophene monolayer is bombarded with circularly polarized light. The electron Hamiltonian's time-reversal symmetry is broken due to the application of circularly polarized light to the 8-borophene monolayer. We artificially characterized the polarized light intensity in terms of amplitude as $AI_0$ and set the initial value to the order of strong light intensity $I = 3 \times 10^{14} Wm^{-2}$, thus inducing a gap of approximately $\Delta \approx 13 meV$ and $\Omega = 2500$ THz [65]. This indicates the incoming beam's energy, which should be significantly more than the bandwidth of borophene.

The Hamiltonian of the $\beta$-Pmmn borophene electron (hole) interacting with surface optical phonon mode can be written as,

$$H = H_\eta^{eff}(\vec{k}) + \sum_{\vec{q},\mu} \hbar\omega_\mu(\vec{q}) b_{\vec{q}}^\dagger b_{\vec{q}} + H_{e-ph} \quad (11)$$

$H_\eta^{eff}(\vec{k})$ is the effective time-independent Floquet Hamiltonian. The second part term represents the free phonon Hamiltonian of the system, and the last term denotes the electron-phonon interaction part, which can be given by

$$H_{e-ph} = \sum_{\vec{q},\mu} [M_{\vec{q},\mu}(z) e^{i\vec{q}\cdot\vec{r}} b_{\vec{q}} + H.C.]. \tag{12}$$

Where the interaction amplitude of electrons with SO phonons of the sub-strate is given by $M_{\vec{q},\mu}(z) = \sqrt{e^2 \kappa \hbar \omega_{\mu,SO}/2\varepsilon_0 q}\, e^{-qz}$, and is used to derive the motion of the interacting with surface optical phonon mode. $\omega_{\mu,SO}$ is the surface optical phonon frequency with two branches ($\mu = 1,2$). $\varepsilon_0$ is the vacuum permittivity, alternatively can be referred to as permittivity of free space. The dielectric constant of the substrate can be written as, $\kappa = (\kappa_0 - \kappa_\infty)/[(\kappa_0 + 1)(\kappa_\infty + 1)]$ which represents the polarizability of the substrate. Here, $\kappa_0$ ($\kappa_\infty$) represents the low (high) frequency the dielectric constants of the subsystem corresponding to the substrate of the crystal. The annihilation (creation) operator $b_{\vec{q}}$ ($b_{\vec{q}}^\dagger$) for SO-phonon with $q$ phonon wave-vector, the operators satisfy the following commutation relation, $[b_i, b_j^\dagger] = \delta_{ij}$. We should specify the dependence of the cut-off wave vector of the optical phonon modes, as $q_{cut-off} = 0.45$ [27]. In the $e - ph$ interaction part, $M_{\vec{q},\mu}(z)$, the values of $z$ is the internal distance between the polar substrates and β-borophene. In this study, we use the values of the distance parameter as approximately, $z = 1Å^{-1}$, and analyze in a constant value. The magnitude of SO-phonon contribution energy for all crystals drastically decreases as $z$ increases initially and then increases gradually with decreasing $z$ values [27]. In practical calculation, to examine its effectiveness, we use $h - BN$ polar material as a substrate, which has the highest surface optical phonon frequency. For a specific substrate, effective screening constant value, dielectric constants, and surface optical phonon mode parameters have been obtained from [70]. The parameters give insight into understanding the strong polaronic effect of materials. With the purpose of creating an all coupling theory for Fermi polarons, in the frame of many-body quantum field theory, Lee-Low-Pines theory constitutes a powerful tool for Fröhlich polarons. The LLP theory contains two double unitary transformations.

$$U_1 = exp\left[-i\vec{r} \cdot \sum_{\vec{q}} \hbar\vec{q}\, b_{\vec{q}}^\dagger b_{\vec{q}}\right] \tag{13a}$$

$$U_2 = exp\left[\sum_{\vec{q},\mu} \left(f_\mu(\vec{q}) b_{\vec{q}}^\dagger - f_\mu^*(\vec{q}) b_{\vec{q}}\right)\right] \tag{13b}$$

The $U_1$ transformation eliminates the $\vec{r}$ electron coordinates in the Hamiltonian. Applying the transformation on $\vec{p}_i$ the electron operators and $b_{\vec{q}}$ phononic operators are given by $\wp_i = U_1^{-1} \vec{p} U_1 = \vec{p}_i - \sum_{\vec{q}} \hbar\vec{q} b_{\vec{q}}^\dagger b_{\vec{q}}$ and $\flat_{\vec{q}} = U_1^{-1} b_{\vec{q}} U_1 = b_{\vec{q}} e^{-i\vec{q}\cdot\vec{r}}$ respectively. Thus, the first transformed form of the Hamiltonian can be represented as, $H_1 = U_1^{-1} H U_1$. The second transformation changes the phonon coordinates, allowing us to account for the dressed electron states caused by the coherent states of the phonon field. By the way of the optical phonon transform has the form, $U_2^{-1} b_{\vec{q}} U_2 = b_{\vec{q}} + f_\mu^*(\vec{q})$. According to the transformation $f_\mu(\vec{q})$ is a variational function. This transformation allows us to consider the states of dressed electrons caused by the phonon field of the substrate, causing polarization in the lattice. Because of the two consecutive unitary transformations, $H_2 = U_2^{-1} U_1^{-1} H U_1 U_2 = U_2^{-1} H_1 U_2$, the new Hamiltonian is performed. It generates coherent states from a phonon vacuum. By using phonon vacuum state $|0\rangle$ on the transformed Hamiltonian. Thus it satisfies $b|0\rangle = 0$, which is chosen on the basis of the zero absolute temperature approximations. The expectation values of the transformed Hamiltonian ($H' = \langle 0|H_2|0\rangle$) for the examined system are as follows:

$$H' = \left(\hbar v_x^{eff}\left[k_x - \sum_{\vec{q},\mu} q_x |f_\mu(\vec{q})|^2\right]\right)\sigma_x + \left(\hbar v_y^{eff}\left[k_y - \sum_{\vec{q},\mu} q_y |f_\mu(\vec{q})|^2\right]\right)\sigma_y$$

$$+ \left(\hbar v_t\left[k_y - \sum_{\vec{q},\mu} q_y |f_\mu(\vec{q})|^2\right]\right)\sigma_0 + \Delta\sigma_z$$

$$+ \left[\sum_\mu \sum_{\vec{q}} \hbar\omega_\mu |f_\mu(\vec{q})|^2 + \sum_\mu \sum_{\vec{q}} \left(M_{\vec{q},\mu}(z)f_\mu(\vec{q}) + M_{\vec{q},\mu}^*(z)f_\mu^*(\vec{q})\right)\right]\sigma_0 \quad (14)$$

By applying the functional minimization procedure [70] $\left(\partial H'/\partial f_\mu(\vec{q}) = 0\right)$ of Eq. (14) the determinant of the derivative of the Hamiltonian is made to vanish, implying that it is extremal under variations of its eigenvalues [71, 72]. Thus, we get the solutions to the variational parameters, which are as follows:

$$f_\mu(\vec{q}) = \frac{-M_{\vec{q},\mu}^*(z)}{\hbar(\omega_\mu - v_t q_y) \pm \sqrt{\left(\hbar v_x^{eff}\right)^2 q_x^2 + \left(\hbar v_y^{eff}\right)^2 q_y^2}}$$

In here, $f_\mu(\vec{q})$ is considered as a variational function. The new variation parameters $\chi_x$ and $\chi_y$ along the $x$ and $y$ directions, respectively. The only preferred direction in the system is the direction of momentum, i.e., $k_x$ and $k_y$, due to the symmetry rules, so $\sum_{\vec{q},\mu} q_i |f_\mu(\vec{q})|^2$ should differ by $k_i$ by a scalar. Thus $\chi_i$ corresponds to $\sum_{\vec{q},\mu} q_i |f_\mu(\vec{q})|^2 / k_i$, that is valid for an intermediate coupling regime. In terms of the new variation parameters Eq.(14) can be written as,

$$H' = \left[\hbar v_x^{eff} k_x (1-\chi_x)\sigma_x + \hbar v_y^{eff} k_y (1-\chi_y)\sigma_y + \hbar v_t k_y (1-\chi_y)\sigma_0\right] + \Delta\sigma_z$$

$$+ \left[\sum_\mu \sum_{\vec{q}} \hbar\omega_\mu |f_\mu(\vec{q})|^2 + \sum_\mu \sum_{\vec{q}} \left(M_{\vec{q},\mu}(z)f_\mu(\vec{q}) + M_{\vec{q},\mu}^*(z)f_\mu^*(\vec{q})\right)\right]\sigma_0 \quad (15)$$

unperturbed effective part energy satisfies the Eq.(5) can be expressed as,

$$E_{eff} = \hbar v_x^{eff} k_x \langle\sigma_x\rangle + \hbar v_y^{eff} k_y \langle\sigma_y\rangle + \hbar v_t k_y \langle\sigma_0\rangle + \Delta\langle\sigma_z\rangle. \quad (16)$$

The perturbed, interaction part is,

$$E_p = E_x \langle\sigma_x\rangle + E_y \langle\sigma_y\rangle + E_t \langle\sigma_0\rangle + E_f \langle\sigma_0\rangle \quad (17)$$

Each part of the Hamiltonian ($\mathcal{E}_x, \mathcal{E}_y, \mathcal{E}_t$, and including $\mathcal{E}_f$) contains summations over $\vec{q}$ in Eq.(17). The summation can be convertible into an integral over phonon wave-vector, $\sum_{\vec{q}} \to (A/4\pi^2) \iint dq_x dq_y$ and $A$ is the surface area of the β-borophene monolayer.

$$\mathcal{E}_x = -(\hbar v_x^{eff})^2 \sum_{\vec{q}\mu} q_x k_x \left( \frac{|M_{\vec{q},\mu}(z)|^2}{\left(\omega + \Omega(q_x, q_y)\right)^2} \right)$$

$$\mathcal{E}_y = -(\hbar v_y^{eff})^2 \sum_{\vec{q}\mu} q_y k_y \left( \frac{|M_{\vec{q},\mu}(z)|^2}{\left(\omega + \Omega(q_x, q_y)\right)^2} \right) \quad (18)$$

$$\mathcal{E}_t = -\hbar v_t \sum_{\vec{q}\mu} q_y \left( \frac{|M_{\vec{q},\mu}(z)|^2}{\left(\omega + \Omega(q_x, q_y)\right)^2} \right)$$

$$\mathcal{E}_{\int.} = \sum_{\vec{q}\mu} \left[ \hbar\omega_\mu \left( \frac{|M_{\vec{q},\mu}(z)|^2}{\left(\omega + \Omega(q_x, q_y)\right)^2} \right) - 2\left( \frac{|M_{\vec{q},\mu}(z)|^2}{\left(\omega + \Omega(q_x, q_y)\right)^2} \right) \right]$$

Here, $\omega = \hbar\omega_\mu - \hbar v_t q_y$ abbreviated as. The expression that $\Omega(q_x, q_y)$ includes $\vec{q}$ dependence in the denominator is defined as the following:

$$\Omega(q_x, q_y) = \sqrt{(\hbar v_x^{eff})^2 q_x^2 + (\hbar v_y^{eff})^2 q_y^2}.$$

In Eq. (17) $\langle \sigma_i \rangle = \langle \Psi | \sigma_i | \Psi \rangle$, $(i = x, y)$ and the normalized wavefunctions are coming from the unperturbed part and Having obtained the time-averaged Hamiltonian

$$\langle \sigma_i \rangle = \frac{v_i^{eff} k_i}{E_{CV}(k_x, k_y)}.$$

Thus, given to the expectation energy eigenvalues are supplied and,

$$E = E_{eff} + E_p$$

its explicit form can be written as,

$$E = \lambda \left( E_{CV}(k_x, k_y) - \mathcal{E}_x - \mathcal{E}_y \right) + \bar{v}_t k_y + \mathcal{E}_{\int.} - E_0$$

where $\lambda = \pm$, modulated bandgap can be denoted as the energy difference below,

$$\Delta E_{gap} = 2 \left( E_{CV}(k_x, k_y) - \mathcal{E}_x - \mathcal{E}_y \right) \quad (19)$$

$$\Delta E_{gap} = 2 \left[ \left( \sqrt{(\hbar v_x^{eff} k_x)^2 + (\hbar v_y^{eff} k_y)^2 + \Delta^2} \right) - \sum_{\vec{q}\mu} q_x k_x \left( \frac{|M_{\vec{q},\mu}(z)|^2}{\left(\omega + \Omega(q_x, q_y)\right)^2} \right) \frac{(\hbar v_x^{eff})^2}{E_{CV}(k_x, k_y)} \right.$$

$$\left. - \sum_{\vec{q}\mu} q_y k_y \left( \frac{|M_{\vec{q},\mu}(z)|^2}{\left(\omega + \Omega(q_x, q_y)\right)^2} \right) \frac{(\hbar v_y^{eff})^2}{E_{CV}(k_x, k_y)} \right]. \quad (20)$$

In this calculation, we have used right-handed for circularly polarized light in $K$ the valley. This is because right-handed circular polarization increases the band gap for the $K$ valley while decreasing it for the $K'$ valley. The reverse effect for left-handed circular polarization. However, the linearly polarized off-resonant light cannot open a gap but can induce an anisotropic band structure for the Dirac materials [65]. Therefore, only one valley is significant for the low-energy electronic features. Besides, the magnitude of band gap modulation in the intermediate coupling regime can be altered by circularly polarized light (CPL), so the magnitude of the band gap is a function of the CPL. From Eq. (19) and (20), one can easily see the polaronic and polarized light contributions on the modulated bandgap of $\beta$-borophene monolayer.

3. Results and Discussion

The study of matter's interaction with light fields, i.e., electromagnetic radiation, is one of the most significant issues in time-dependent quantum mechanics. In this study, the effective low-energy Dirac equation for the electron states in the presence of light-matter interactions, which are the result of circularly polarized light (CPL) colliding with charged particles and the surface optical phonon contribution to the monolayer β-borophene. Fig.1 shows the schematic drawing of the borophene lattice in three dimensions and a top view of the optimized crystal structure of borophene in a unit cell. Fig. 2 shows the band gap examination of the circularly polarized light and surface optical phonon contribution on the monolayer β-borophene. First panel from the left part of Fig.2 evaluates the monolayer $\beta$-borophene which is represented by a Dirac cone for electrons and holes. The Dirac forms touch each other at $k = 0$ similar to the graphene. However, in this case, borophene has an anisotropic structure due to distort along the $k_y$ axis. In this study can be seen from the figure that we investigate the band gap emerging state in a constant each $k$, i.e., $k_x = k_y = 0.001 Å^{-1}$. This promising structure becomes a semiconductor in even the very small k values and its calculated band gap energy eigenvalue is $14.4 meV$. The second panel (green) from the left part of Fig.2 shows a borophene monolayer which is illuminated by off-resonant light of amplitude which represents the $3 \times 10^{12} W cm^{-2}$ intensity of polarized light. The effect of photon absorption at the degenerate Dirac points is to open a gap with a magnitude of $\Delta = \frac{\xi 2e^2 I v_x v_y}{c\varepsilon_0 \hbar \Omega^3}$. Polarized off-resonant light induces a $29.6 meV$ band gap energy at a small $k_x = k_y = 0.001 Å^{-1}$ value.

The highest-frequency phonon modes at the $\Gamma$-point in two-dimensional borophene open a dynamic band gap similar to the graphene. The electron-phonon interaction increases the renormalized mass of the polaron because, the renormalized mass of the polaron reflects the mass of the quasi-particle that is generated by the electron, which induces the lattice distortion. In this study, we adapted monolayer borophene and boron is quite a light element. But, actually, the electron drags in the heavy ion cores along with it, giving the impression that it has a larger mass [73]. The internal distance between the polar substrates and β-borophene, which is z value. The distance is increase as z rises. The renormalized mass at the crystal surface reduces dramatically because of the influence given by the electron-SO phonon interaction, which diminishes fast as z increases [27]. In the third (red) panel from the left of Fig. 2, one can see the only polaronic contribution to the β-Borophene that the internal distance between the polar substrates and β-borophene is constant 1 $(Å^{-1})$. SO-phonon contribution on the monolayer structure has used $h - BN$ substrate that has the smallest effective screening constant [27]. This gives the smallest surface optical phonon energy contribution [73] on the structure, which is almost $142 meV$. It is also seen from the panels that, when we consider the effects separately, the polaronic effect is more effective than the CPLight effect at the small values of $k$. Then we come to the fourth panel (blue) of Fig.2, which represents the combined effects of CPLight and SO-phonon contributions on $h - BN$ the substrate. This combined effect does not directly contribute as much as the sum of the effects. It even appears to decrease the combined effect of polaronic and CPLight contribution to the monolayer structures such that the band-gap contribution is almost $44 meV$. Finally, in the last part of the figure shows that the enhancement in the energy eigenvalues due to interaction between CPL and SO-phonon interactions decrease with increasing $k$ (blue lines). Besides, with increasing $k$ value the energy contribution vanished entirely at the value of $0.0034 Å^{-1}$. Afterwards, it continued to increase its efficiency with an increasing $k$ value. The other lines (red) represent the only polaronic effect contribution in the energy eigenvalues. It is clear from the figure that the enhancement in the energy eigenvalues due to the polaronic effect decreases with increasing $k$ than the effect contributions completely disappeared at the value of $0.0042\text{Å}^{-1}$. In here, dashed lines show the behavior in the $k_y$ component of the band gap energy, while the straight line shows the behavior in the $k_x$ component of the band gap energy. As also seen from the figure that the anisotropic behavior is better exhibited at high $k$ values. $k_y$ component of the behavior, i.e., dashed lines give the small band gap emerging than the $k_x$ component, i.e., straight lines of the band gap emerging value.

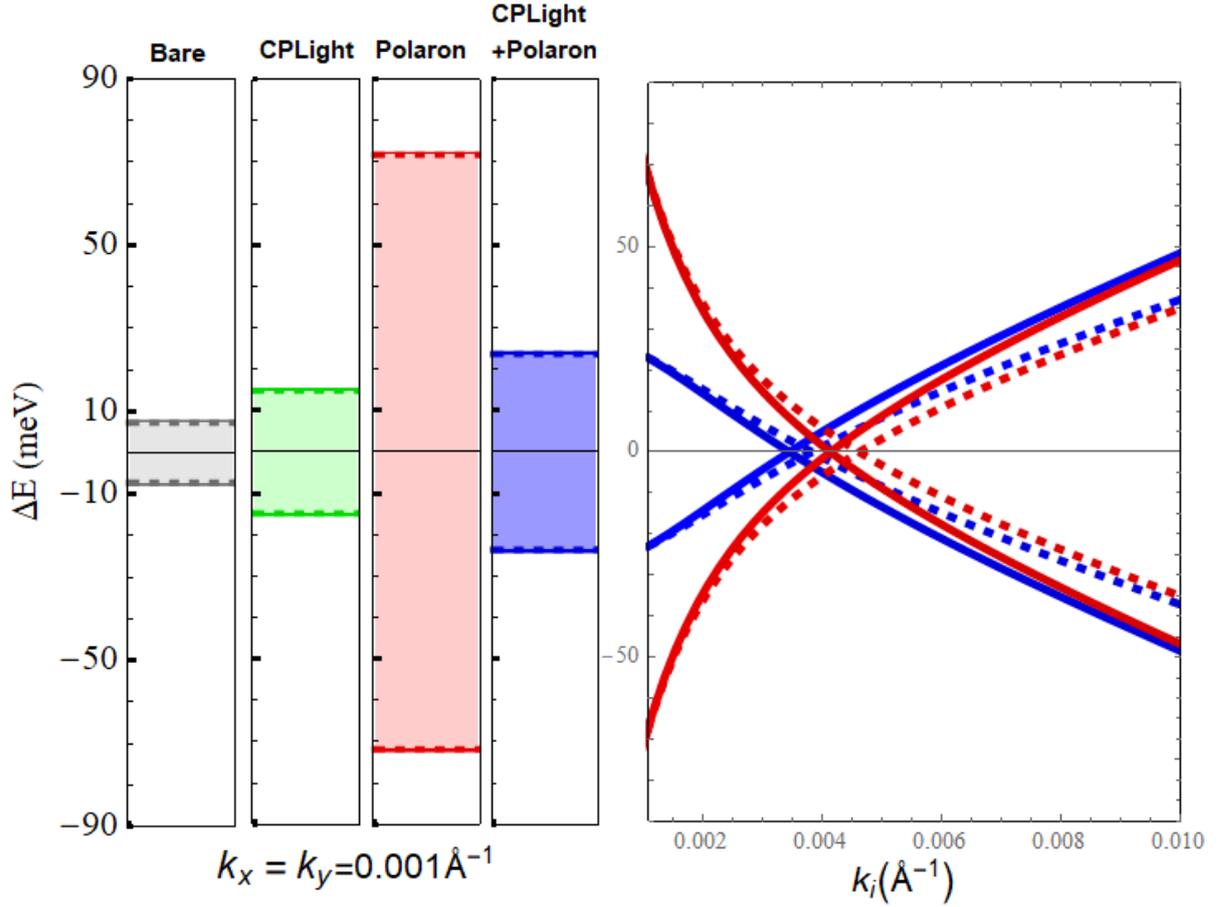

**Fig.2** Bandgap examination of the circularly polarized light and surface optical phonon contribution on the monolayer $\beta$-borophene calculated energies from equation (20) at a constant value of $k_x = k_y = 0.001\text{Å}^{-1}$. The first left panel (grey) represents the tilted monolayer structure that canceled CPL and phonon contribution to it. The second panel (green) is evaluated in the presence of CPL. The third left panel (red) is representative of the polaron self-energy on the gapped monolayer borophene and the fourth panel (blue) shows the combined effects of CPL and SO-phonon contribution. The last part of the figure shows the treatment in the energy eigenvalues with increace $k$.

$k$-dependent behavior of the figure can be seen as a three-dimensional (3D) in Fig. 3. The figure is a useful illustration for understanding the anisotropic attitude of $\beta$-borophene. Likewise, Fig.4 (a) also supports this result. The bottom curve (red line) represents the electron-photon contribution to the β-borophene. As expected, CPL enhancement increases its effectiveness by increasing the $\Delta$ value. However, the increased $\Delta$ values could not enhance the effectiveness of the photonic and phononic contributions (blue line) of the monolayers. Fig.4 (b) shows the intensity of circularly polarized light has an effect on the manipulated band gap. The driving photonic frequency has a strong dependence on the effective energy term describing the effects of circularly polarized light. Increasing photonic frequency decreased the electronic band gap contribution because the intensity of CPL and photonic frequency is inversely proportional. In this study, we only adapted to CLP. Napitu [74] reported that if borophene is subjected to linearly polarized light, its electrical characteristics resemble those of pure borophene in the absence of light. This is particularly relevant to the low-energy dependency of the Dirac cone and the topology of the band structure, as defined by the Berry phase. The electron–photon coupling is equivalent to renormalizing the Fermi velocity and momentum wave-vector in such a way that it changes the slope of the density of states or the spectral weight of longitudinal conductivity.

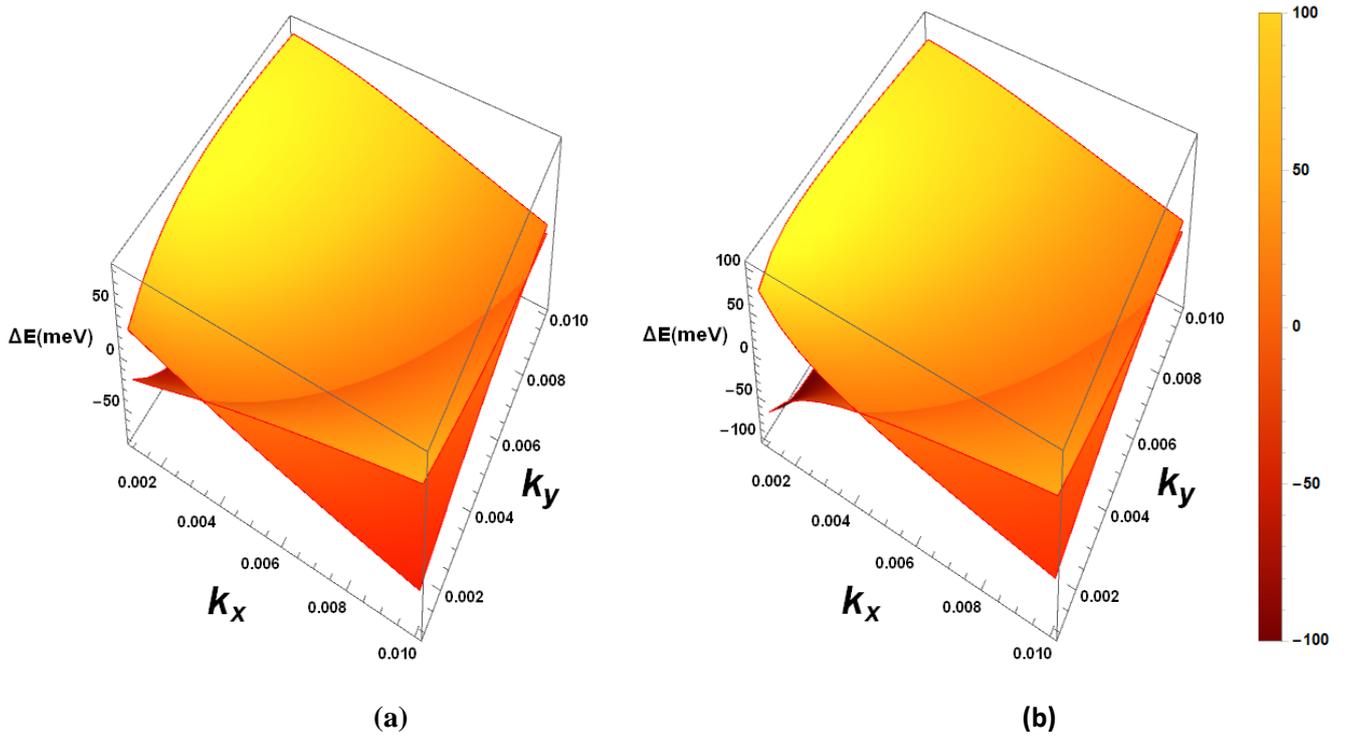

**Fig.3 (a)** band gap examination of the circularly polarized light and surface optical phonon contribution on the monolayer $\beta$-borophene in 3D. **(b)** Polaron self-energy contribution on the monolayer borophene in 3D.

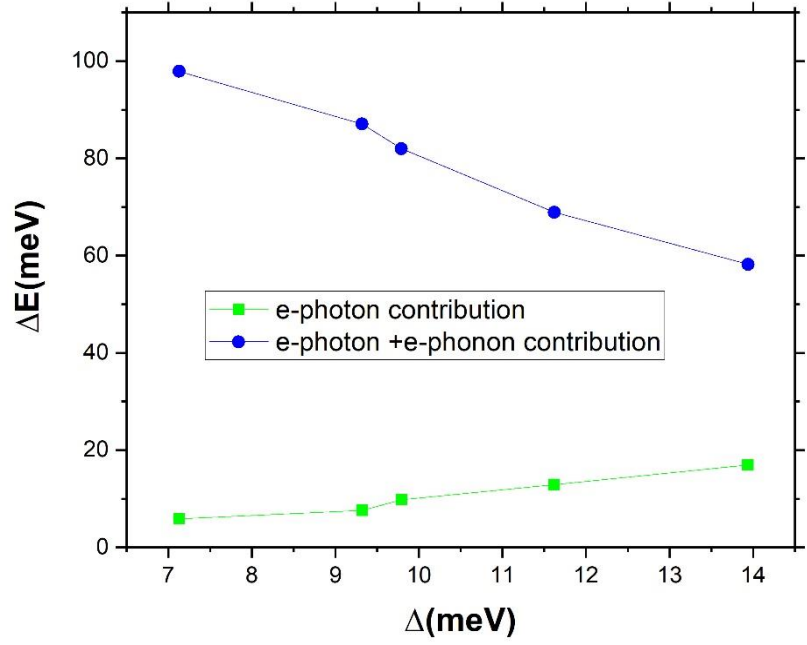

(a)

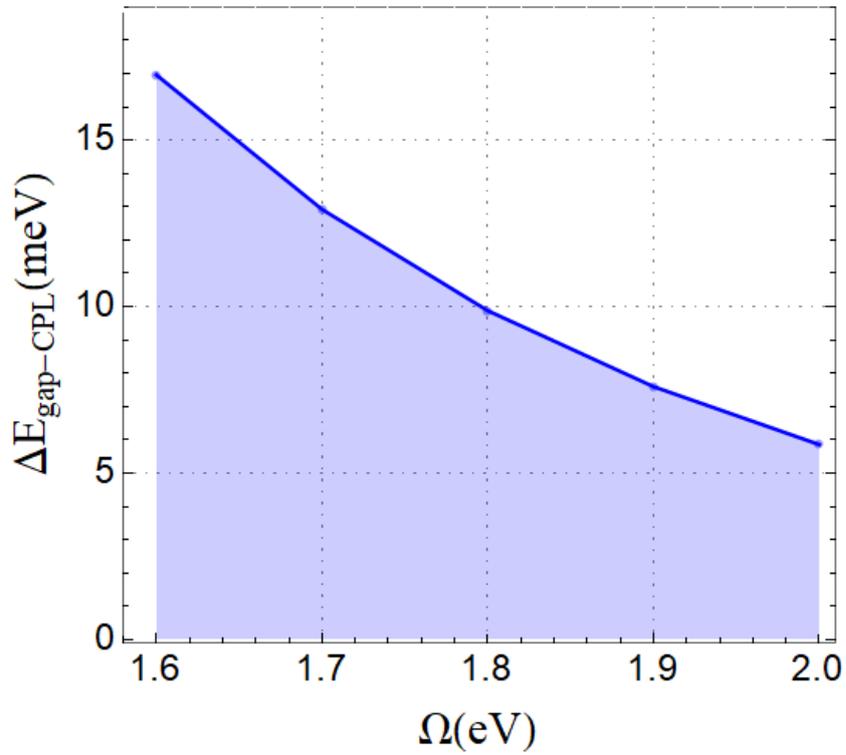

(b)

**Fig. 4 (a)** red dotted line represents to only e-photon contribution, blue dotted line be representative of combined affects of e-photon and e-phonon contribution. **(b)** shows the intensity of circularly polarized light how effect the manipulated band gap.

## 4. Conclusion

In summary, we theoretically studied the electronic properties of the monolayer β-borophene which is irradiated with CPL and dressed with SO-phonon on an h-BN polar substrate. In this model, the brophene uses a continuous model with two $\pi$ bands, where we first neglect the electron-electron interaction. The electron–light matter (photon) coupling is equivalent to renormalizing the Fermi velocity and shows itself as a photo-induced mass term, which gapping manifests as an anomalous Hall effect-driven current [68,69]. The current is defined as the intensity of circularly polarized light that affects the manipulated band gap. Without the electron-light matter coupling, SO phonon contributions have been tested in the material [27]. Our analytical results show that the electron-SO phonon contribution is more effective than the electron-photon interaction, when we examine the contributions of both separately. Moreover, when tested for their combined effect, they did not support each other. Conversely, the combination of the two effects resulted in a smaller electronic band gap than the phonon-dressed effect. An important aspect worth noticing is that the impact of interactions in the monolayer borophene can manipulate its band gap and serve as a band-gap engine for the photonics, energy transportation, and electronic devices.